# All-polarization-maintaining linear cavity fiber lasers mode-locked by nonlinear polarization evolution in stretched pulse regime

Xuanyi Liu, Feng Ye, Minghe Zhao, Boris A. Malomed, H. Y. Fu, and Qian Li

*Abstract*—Nonlinear polarization evolution (NPE) is among the most advanced techniques for obtaining ultrashort pulses with excellent optical performance. However, it is challenging to design environmentally stable NPE fiber oscillators using only polarization-maintaining (PM) fibers. Here, we use the same PM fiber and non-reciprocal phase shifter to design two different devices, which are capable of acting as effective NPE saturable absorbers (SAs) in two all-PM linear cavity fiber lasers. These two laser setups differ in the position of the non-reciprocal phase shifter, the presence of which is crucial for the proposed fiber lasers to reduce their mode-locking thresholds and achieve high repetition rates above 100 MHz. The mode-locking principle and pulse evolution in the laser cavity are investigated theoretically. The first all-PM fiber oscillator emits sub-200 fs stretched pulses with low peak powers. The second oscillator, with a simpler architecture, directly delivers stretched pulses with high peak powers, the spectral bandwidth greater than 30 nm, and the pulse duration less than 90 fs. To the best of our knowledge, 79 fs achieved in this design is the shortest pulse duration provided by PM fiber lasers using NPE mode-lockers.

## I. Introduction

ULTRAFAST lasers are fundamental building blocks of diverse scientific and industrial applications such as terahertz-wave generation [1], optical frequency comb sources [2], and high-efficiency micromachining [3]. Compared to other types of pulsed lasers, fiber lasers are becoming increasingly competitive due to their compactness, low implementation cost, and reduced thermal effects [4]. A key issue in the work with fiber lasers is how to design a laser cavity delivering ultrashort pulses with specific optical properties, such as large spectral bandwidth, short pulse duration, and high repetition rate. Furthermore, maintaining the long-term environmental stability and reproducibility of a mode-locked state is another challenging issue. To handle these issues and develop high-quality and stable pulsed fiber oscillators, robust pulse-picking and delicate pulse-shaping schemes are requisite.

Previously, various saturable absorbers (SAs) have been developed for the use in mode-locked fiber lasers to achieve pulse-picking operation. Incorporating real SAs, such as semiconductor saturable absorber mirror (SESAM) [5], carbon nanotube (CNT) [6], graphene [7], and other two-dimensional (2D) materials into the laser cavity offers a convenient method for fabricating easily self-starting fiber lasers. Nevertheless, the potential shortcomings of SESAMs, such as complicated manufacturing procedures and narrow bandwidth, limit their use. 2D nanomaterials-based fiber lasers suffer from low damage thresholds and performance degradation over time [8]. Nonlinear polarization evolution (NPE) is an effective artificial SA for building high-performance fiber lasers, whose mode-locking mechanisms are based on nonlinear optical effects. In traditional NPE-based mode-locked fiber lasers, the Kerr nonlinearity in weakly birefringent fibers induces the intensity-dependent nonlinear phase shift, which converts the change in polarization state into amplitude modulation. The high-intensity pulses experience lower loss, which acts as a fast SA supporting passive mode-locking. Although NPE mode-locked fiber lasers have achieved superior optical performance, such as high single-pulse energy [9], short pulse duration [10], and high repetition rate [11], the utilization of non-polarization-maintaining (non-PM) fibers restricts their working scenarios and commercialization. In complex environments, where factors such as temperature and vibration may vary dramatically, the laser is likely to lose its mode-locked state, and the optical characteristics may not be maintained. An upgraded version of the traditional NPE technique is to implement NPE mode-locking in PM fibers. All-PM ring-cavity and linear-cavity fiber lasers based on NPE have been proposed and demonstrated by several groups [12]–[26]. To eliminate the group-velocity mismatch (GVM), the length and

This work was supported in part by Overseas Research Cooperation Fund of Tsinghua Shenzhen International Graduate School (HW2020006), in part by Shenzhen Fundamental Research Program (GXWD20201231165807007-20200827130534001), in part by Youth Science and Technology Innovation Talent of Guangdong Province (2019TQ05X227), and in part by Israel Science Foundation (grant No. 1695/22). (*Corresponding authors: H. Y. Fu; Qian Li.*)

Xuanyi Liu and H. Y. Fu are with the Tsinghua Shenzhen International Graduate School and Tsinghua-Berkeley Shenzhen Institute, Tsinghua University, Shenzhen 518055, China (e-mail: hyfu@sz.tsinghua.edu.cn).

Feng Ye, Minghe Zhao, and Qian Li are with the School of Electronic and Computer Engineering, Peking University, Shenzhen 518055, China (e-mail: liqian@pkusz.edu.cn).

Boris A. Malomed is with the Department of Physical Electronics, School of Electrical Engineering, Faculty of Engineering, and Center for Light-Matter Interaction, Tel Aviv University, Tel Aviv 69978, Israel; and the Instituto de Alta Investigación, Universidad de Tarapacá, Casilla 7D, Arica, Chile (e-mail: malomed@tauex.tau.ac.il).

splicing angle of the PM fiber pieces should be accurately adjusted in the all-PM ring cavity fiber laser mode-locked by means of NPE. Off-axis fusion splicings between each PM fiber are required to introduce SA effects into the laser cavity, which increases the complexity of the laser fabrication and impairs the stability of output characteristics. Compared to ring cavity fiber lasers, the GVM caused by strong birefringence in linear cavity fiber lasers can be fully canceled by the Faraday mirror placed at the end of the laser cavity. In 2007, C. K. Nielsen *et al*. have designed an all-PM linear cavity fiber laser emitting 5.6-ps ultrashort pulses at a repetition rate of 5.96 MHz [12]. S. Boivinet *et al*. have demonstrated the realization of different pulsed regimes in a 948-kHz PM fiber laser mode-locked by means of NPE [13]. More recently, numerical modeling and experimental investigation have been performed in a Yb:fiber laser with a linear cavity, which generates dispersion-managed soliton pulses with a spectral bandwidth of 37.84 nm [26]. However, the fundamental repetition rate of these linear cavity fiber lasers is below 10 MHz because long PM fibers are needed to accumulate sufficient nonlinear phase bias. The non-reciprocal phase shifter formed by the combination of a Faraday rotator and wave plates is able to shift the reflectivity curve of SA to effectively reduce the mode-locking threshold. By introducing the non-reciprocal phase shifter into the free-space linear arm [27] or fiber loop [28], the figure-9 fiber laser mode-locked by nonlinear amplifying loop mirror (NALM) exhibits extraordinary self-starting ability of mode-locking. The insertion of the non-reciprocal phase shifter prods the laser towards producing a continuous-wave (CW) output, which facilitates the generation of high-repetition-rate and high-energy pulses [29], [30]. Furthermore, a self-referenced optical frequency comb with superior stability has been implemented in a NALM-based fiber laser [31]. The optical performance of fiber lasers using a phase-biased NALM can be significantly enhanced, and this technique has also been applied to all-PM fiber lasers with linear configurations. Recently, linear-cavity fiber oscillators with a non-reciprocal phase shifter have been intensively investigated [15]–[18]. In particular, our group has simplified the structure of an all-PM linear-cavity fiber oscillator, achieving the highest available (to the best of our knowledge) repetition rate, 133 MHz [18]. The laser operates at anomalous dispersion, and no intra-cavity dispersion engineering has been used. The achievable pulse duration and spectral bandwidth are limited in the soliton regime. To promote the application range of this laser as a broad-spectrum light source, it is very relevant to explore other pulsed regimes, such as ones producing stretched pulses. The implementation of the dispersion-engineering technique allows significant compression of the generated pulses. Besides, although the optical characteristics of linear-cavity fiber lasers mode-locked by means of NPE have been experimentally investigated in Ref. [16]–[18], the influence of the position of the non-reciprocal phase shifter on the output characteristics is not known. Another motivation to study such fiber lasers is to simplify and optimize their laser architectures to reduce the manufacturing costs and enhance the optical performance.

In this paper, two novel linear-cavity fiber oscillators with a simple design, using different locations of the non-reciprocal phase shifter, are presented. In these setups, the NPE mode-locking mechanism is implemented in all-PM fibers, enhancing the environmental stability of the output characteristics. Both laser configurations operate in the stretched pulse regime, the generated pulses are characterized by a large spectral bandwidth, short pulse duration, and high repetition rate. The first configuration delivers sub-200 fs stretched pulses at the fundamental repetition rate of 114.97 MHz. In contrast, sub-90 fs high-peak-power pulses at 116.76-MHz repetition rate are generated by the second configuration. The phase and amplitude noise of the proposed all-PM fiber lasers are measured and analyzed in the free-running state. The combination of excellent optical performance, low-noise operation, and environmental stability makes these linear cavity fiber lasers appropriate for a wide range of applications.

## II. LASER ARCHITECTURES AND THEORETICAL ANALYSIS

Two experimental laser architectures (Laser1 and Laser2) that differ by the position of the non-reciprocal phase shifter are schematically illustrated in Figs. 1(a-b). Highly reflective mirrors (M1 and M2) at both ends of the laser cavity lend the PM fiber oscillator a rectilinear structure. Both fiber oscillators are composed of two free-space portions and a segment of the PM fiber. The latter one is spliced by a PM erbium-doped fiber (Liekki Er80-4/125-HD-PM, PM-EDF) with normal group velocity dispersion (GVD) 28.0 ps$^2$/km and length 33 cm, and a PM single-mode fiber (PM-SMF) with anomalous GVD −23.0 ps$^2$/km at 1550 nm and length 51 cm. The net cavity dispersion is ≈ −0.0025 ps$^2$, hence the laser operates in the nearly-zero-GVD mode, which supports the generation of stretched pulses. The active fiber is pumped by a single-mode semiconductor laser diode (LD) with a maximum pump power of 630 mW. The Semi-wavelength division multiplexer (Semi-WDM) employed in the laser cavity has a dual function [18]. The first function is to inject the 976 nm pump light into the gain medium. Secondly, it serves as a collimator. In Laser1, the non-reciprocal phase shifter comprising a Faraday rotator (FR) and a λ/8-wave plate (EWP) is inserted between the polarization beam splitter (PBS) and half-wave plate (HWP), as shown in the encircled box of Fig. 1(a). Differently, the EWP of Laser2 is installed between the FR and M1, i.e., the non-reciprocal phase shifter is moved to the opposite end of the PM fiber, as shown in the encircled box of Fig. 1(b). Compared to Laser1, only one FR is required in Laser2, which greatly simplifies the rectilinear structure of the laser and reduces the manufacturing cost. The PBS extracts linearly polarized pulses from the laser cavity.

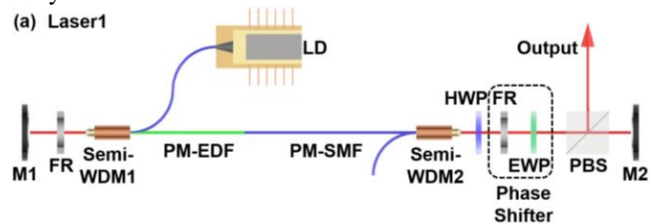

(a) Laser1

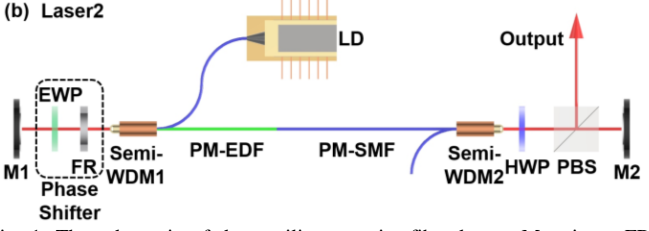

Fig. 1. The schematic of the rectilinear cavity fiber lasers. M: mirror; FR: Faraday rotator; Semi-WDM: Semi-wavelength division multiplexer; PM-EDF: polarization-maintaining erbium-doped fiber; PM-SMF: polarization-maintaining single-mode fiber; HWP: half-wave plate; EWP: λ/8-wave plate; PBS: polarization beam splitter; LD: laser diode.

In the all-PM laser cavities presented above, effective artificial SAs play the crucial role in supporting the NPE mode-locking mechanism. Figure 2(a) depicts the artificial SA of Laser1 and the evolution of the pulse polarization states in the three-dimensional space. The artificial SA consists of several free-space elements for polarization controlling and a segment of PM fiber serving as the Kerr medium. The mode-locking mechanism relies on the nonlinear propagation of orthogonally polarized pulses with unequal intensity along the fast and slow axes of the PM fiber. The non-reciprocal phase shifter incorporated into the artificial SA plays a significant role in the linear cavity fiber laser. The function of the shifter is to modify the reflectivity, so that its nonzero value and a non-vanishing slope at zero phase difference are provided, to maintain appropriate self-starting capabilities and reduce the mode-locking threshold for the rectilinear cavity fiber laser [15]–[18]. Another essential point is that the combination of a 45-degree rotation FR and the end mirror provides the 90-degree rotation of the orthogonally polarized pulses. Consequently, when the reflected orthogonal pulses pass through the PM fiber again, the GVM effect introduced by the strong birefringence of the PM fiber can be fully removed. The artificial SA of Laser2 with a simpler structure is presented in Fig. 2(b).

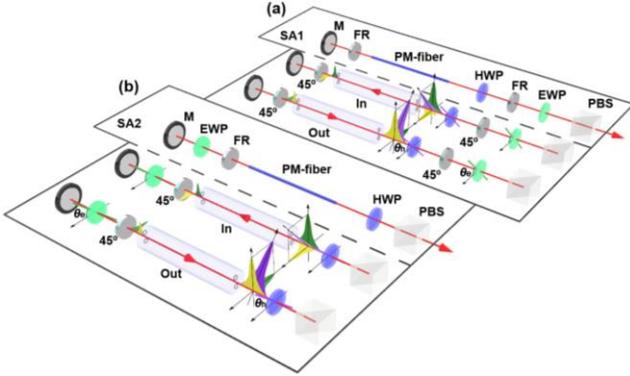

Fig. 2. Artificial SA structures and the evolution of the pulse polarization states in three-dimensional space. M: mirror; FR: Faraday rotator; PM-fiber: polarization-maintaining fiber; HWP: half-wave plate; EWP: λ/8-wave plate; PBS: polarization beam splitter.

Although the two SAs shown in Figs. 2(a-b) have similar structures, it is expected that the difference in the position of the non-reciprocal phase shifter has a great influence on the modulation depth, saturable losses, modulation period, *etc*. Because in SA1, the orthogonally polarized pulses are phase-biased after two passes through the PM fiber, while in SA2, they are phase-biased with only a single pass through the PM fiber. To figure out the difference between the two SAs and gain deeper understanding of the pulse-picking mechanism of each SA, a theoretical calculation of the artificial SA reflectivity, as a function of the nonlinear phase shift, should be performed. The effect of each optical element on the polarized pulse can be represented by Jones matrices as a linear transformation [32]. The matrices for each optical device shown in Fig. 2 are given in Table 1. The nonlinear phase shift for the orthogonal pulses double-propagating through the PM fiber is defined as $\Delta\varphi_{nl}$, which is mainly contributed to by the self-phase modulation (SPM) and cross-phase modulation (XPM) [21]. The accumulation of the nonlinear phase shift follows the relation: $\Delta\varphi_{nl} = \Delta\varphi_{nl1} + \Delta\varphi_{nl2}$, where $\Delta\varphi_{nl1}$ and $\Delta\varphi_{nl2}$ represent the nonlinear phase shift of the pulses entering and leaving the PM fiber, respectively.

TABLE I
JONES MATRICES FOR THE OPTICAL ELEMENTS

| Optical elements | Jones matrices |
|---|---|
| Half-wave plate | $M_{HWP}(\theta_h) = \begin{pmatrix} \cos\theta_h & -\sin\theta_h \\ \sin\theta_h & \cos\theta_h \end{pmatrix} \begin{pmatrix} e^{-\frac{i\pi}{2}} & 0 \\ 0 & e^{\frac{i\pi}{2}} \end{pmatrix} \begin{pmatrix} \cos\theta_h & \sin\theta_h \\ -\sin\theta_h & \cos\theta_h \end{pmatrix}$ |
| λ/8-wave plate | $M_{EWP}(\theta_e) = \begin{pmatrix} \cos\theta_e & -\sin\theta_e \\ \sin\theta_e & \cos\theta_e \end{pmatrix} \begin{pmatrix} e^{-\frac{i\pi}{8}} & 0 \\ 0 & e^{\frac{i\pi}{8}} \end{pmatrix} \begin{pmatrix} \cos\theta_e & \sin\theta_e \\ -\sin\theta_e & \cos\theta_e \end{pmatrix}$ |
| Faraday rotator (45°) | $M_{FR} = \begin{pmatrix} \frac{\sqrt{2}}{2} & \frac{\sqrt{2}}{2} \\ -\frac{\sqrt{2}}{2} & \frac{\sqrt{2}}{2} \end{pmatrix}$ |
| Polarization beam splitter | $M_{PBS,trans} = \begin{pmatrix} 1 & 0 \\ 0 & 0 \end{pmatrix}$ |
| Mirror | $M_M = \begin{pmatrix} -1 & 0 \\ 0 & -1 \end{pmatrix}$ |
| Nonlinear phase shift | $M_{NPL1} = \begin{pmatrix} e^{i\Delta\varphi_{nl1}} & 0 \\ 0 & 1 \end{pmatrix}$, $M_{NPL2} = \begin{pmatrix} 1 & 0 \\ 0 & e^{i\Delta\varphi_{nl2}} \end{pmatrix}$ |

The PBS is set as the starting point of the incident pulse. After passing through the PBS, the polarization state of the incident pulse can be expressed by a normalized field vector $e_x = (1, 0)$. The subsequent transmission of the pulses through the polarization-controlling optics and PM fiber is represented by simple matrix multiplication. As a result, the transformation of the pulses' electric field, $E_{trans1} = (E_{x1}, E_{y1})$, in the course of the roundtrip can be written as

$$E_{trans1} = M_{PBS,trans}M_M M_{PBS,trans}M_{EWP}(\theta_e)M_{FR}M_{HWP}(\theta_h)M_{NLP2}M_{FR}M_M \\ M_{FR}M_{NLP1}M_{HWP}(\theta_h)M_{FR}M_{EWP}(\theta_e)e_x, \quad (1)$$

for SA1, where parameters $\theta_h$ and $\theta_e$ represent the rotation angle of the HWP and EWP, respectively. For SA2 with the simplified structure, the transformed electric field $E_{trans2} = (E_{x2}, E_{y2})$ can be expressed in the simpler form, as

$$E_{trans2} = M_{PBS,trans}M_M M_{PBS,trans}M_{HWP}(\theta_h)M_{NLP2}M_{FR}M_{EWP}(\theta_e)M_M \\ M_{EWP}(\theta_e)M_{FR}M_{NLP1}M_{HWP}(\theta_h)e_x. \quad (2)$$

Thus, the corresponding reflectivity function for each SA can be calculated as follows,

$$R_{1,2} = |E_{x1,2}|^2 \quad (3)$$

It follows from Eq. (3) that the change of the reflectivity curve is determined by the rotation angle of the wave plates ($\theta_h$ and $\theta_e$) and the nonlinear phase shift ($\Delta\varphi_{nl}$). Figures 3(a-d) show the accordingly calculated normalized reflectivity of SA1 as a function of the nonlinear phase shift for different wave-plate rotations. Note that if the value of the reflectivity function at zero nonlinear phase shift has a nonzero value and, more importantly, it exhibits a non-vanishing slope, the build-up of pulses from noise at low power is facilitated [33]. Operation points marked by hollow black circles in all the figures represent the position on the curve where the phase bias is applied to the artificial SA. When the rotation angle of the EWP is adjusted to 0º, 15º, 30º, and 45º, the respective number of the reflectivity curves with phase biases shows a trend to the increase, viz., 2, 3, 4, and 6, respectively. Figures 3(a-c) demonstrate the cyclicity and symmetry of the reflectivity curves. The blue and red curves corresponding to HWP rotation angles of 0º and 45º, respectively, are identical and feature no periodic modulation, indicating that the reflectivity curve is cycled with the 45º rotation of the HWP as a cycle. Curves of different colors in Fig. 3 are mirror-symmetrically distributed around the cyan one. In Figs. 3(a-c), most reflectivity curves have a modulation period of $4\pi$, while the other cyan curve, representing the HWP rotation angle of 22.5º, has a modulation period of $2\pi$. Figure 3(d) shows the case where the rotation angle of the EWP is adjusted to 45º, that is, after the orthogonally polarized pulses are rotated 45º by the FR, their polarization states correspond to the fast and slow axes of the EWP. In this case, changing the angle of the HWP affects only the modulation depth at a fixed phase bias ($\pi/2$) and the modulation period ($2\pi$).

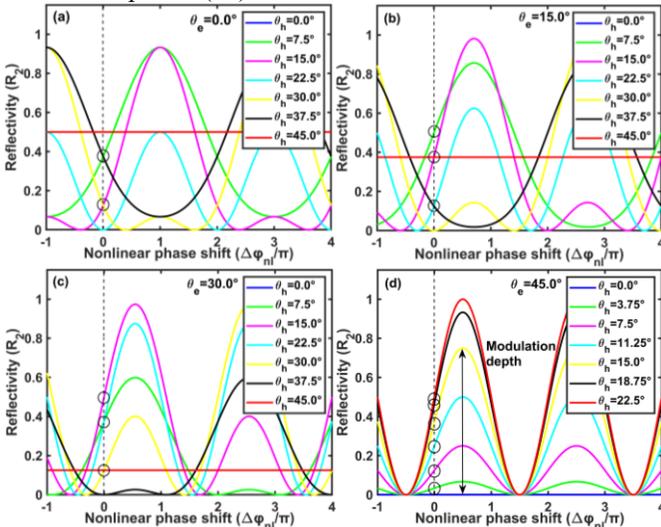

Fig. 3. The normalized reflectivity ($R_1$) as a function of nonlinear phase shift for different wave-plate angles, $\theta_h$ and $\theta_e$. (a)-(c) $\theta_e = 0°$, 15º, and 30º. Tuning angle $\theta_h$ affects the modulation depth, phase bias, and modulation period. (d) In the case of $\theta_e = 45°$, tuning angle $\theta_h$ affects only the modulation depth, with a fixed nonlinear phase shift of $\pi/2$.

In comparison to SA1, setup SA2 exhibits very different reflectivity properties, as shown in Figs. 4(a-d). In Fig. 4(a), the rotation angle of the EWP is set to 0º, while the rotation angle of the HWP increases from 0º to 22.5º in steps of 3.75º. It is found that the corresponding reflectivity curve exhibits the modulation depth decreasing from 100% to 0%, and no phase bias is observed. The red curve at $\theta_h = 22.5°$ indicates no reflectivity modulation, the pulses being totally reflected. To adjust the value and slope of the reflectivity curve at zero phase difference, the EWP rotating angle is switched to 15º, 30º, and 45º. The respective changes in the reflectivity are obtained by altering the HWP rotating angle from 0º to 45º, as shown in Figs. 4(b-d). At a fixed rotation angle of EWP, the modulation depth of the SA decreases with an increase in $\theta_h$ from 0º to 22.5º. An extreme example, shown in Fig. 4(d), is the case where the rotation angles of the EWP and HWP are both 45º, the modulation depth attains 100%, and the nonlinear phase shift is exactly $\pi/2$, leading to a maximum slope of the entire reflectivity curve. It is seen from the above analysis that any desired modulation depth and phase bias can be obtained by adjusting the rotation angle of HWP and EWP, corroborating that the so-designed SA possesses the capability of reliable and reproducible pulse-picking.

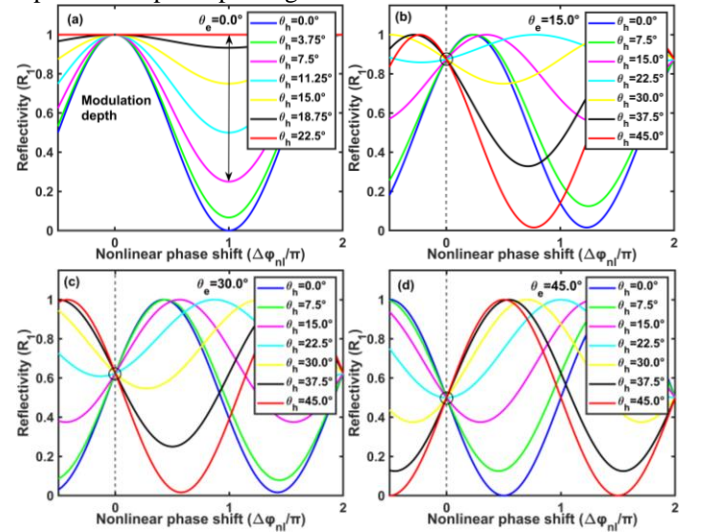

Fig. 4. The normalized reflectivity ($R_2$) as a function of nonlinear phase shift for different wave-plate angles, $\theta_h$ and $\theta_e$. (a) At $\theta_e = 0°$, tuning angle $\theta_h$ changes solely the modulation depth, but not the phase bias; (b)-(d) $\theta_e = 15°$, 30º, and 45º. In these cases, tuning angle $\theta_h$ affects both the modulation depth and phase bias.

III. EXPERIMENTAL RESULTS

A. Characteristics of Pulses Produced by Mode-locked Laser1

Self-starting mode-locking can be achieved by appropriately fine-tuning the positions of the two wave plates and raising the pump power to a sufficiently high value. At high powers, spectra with CW components or multi-pulsing emerge. Gradually reducing the pump power is required to maintain stable emission of clean pulses. At the mode-locking threshold of 530 mW, Laser1 switches from the CW operation to the pulsed mode. When the pump power is decreased from 280 mW to 200 mW, the average output power of the generated individual pulse drops from 7.71 mW to 1.56 mW. The spectral and temporal diagnostics of the generated pulses measured at the pump power of 280 mW are produced in Fig. 5. As shown in Fig. 5(a), the spectrum measured by an optical spectrum analyzer (AQ6370D Yokogawa) is highly structured and features several intensity spikes in the central part. The sidebands on both sides of the spectrum are inconsistent with the typical spectral features of stretched pulses. The PM fiber can be designed as a Lyot filter based on the birefringence effect. By adjusting the polarization state of the pulse and the length of the PM fiber, the modulation depth and spectral modulation interval of the filter are tunable [34]. This principle is utilized to realize dual-wavelength mode-locking in a figure-9 PM fiber laser [35]. Furthermore, the insertion of PM fibers into ultrafast lasers can introduce strong birefringence-induced phase-matching, which affects the temporal and spectral characteristics of the generated pulses [36], [37]. Therefore, the NPE mode-locking mechanism implemented in PM fibers induces the observed spectral sidebands. The pulse's spectral full width at half-maximum (FWHM) is 32.9 nm, which is broader than the conventional soliton spectrum. A radio-frequency (RF) signal analyzer (N9030B, Agilent) with a bandwidth from 3 to 50 GHz is utilized to record the fundamental RF spectrum and its harmonics, presented in Fig. 5(b). The fundamental frequency peak centered at 114.967 MHz exhibits a high signal-to-noise ratio (SNR) of 73.5 dB with a resolution bandwidth of 10 kHz. The inset of Fig. 5(b) shows the harmonic RF spectra with a span range of 1 GHz, demonstrating good mode-locking performance and stable single-pulse operation. The temporal traces of mode-locked pulses are explored by an oscilloscope (Keysight, DSO-X 6004A) with the help of a 3 GHz bandwidth InGaAs photodetector. As illustrated in Fig. 5(c), the pulse interval is 8.70 ns, which is determined by the pulse propagating forth and back in the PM fiber and free space of the linear laser cavity. The oscilloscope screenshot in an expanded time window (20 ms/div) indicates that the pulse train is free of Q-switched modulation. The output pulse is directly injected from the PBS into a commercial autocorrelator (APE PulseCheck, 150) to measure autocorrelation traces. Fig. 5(d) shows the fringe-resolved and intensity autocorrelation (inset) traces. The pulse duration is estimated to be 157 fs, assuming the sech pulse's profile. Two parasitic sidelobes appear on both sides of the autocorrelation trace, which is mainly caused by the highly structured spectrum. The interval between fringes shown in Fig. 5 (d) is 0.78 ps, corresponding to the Fourier transform of the spectral spacing of 10.3 nm between the intensity spikes [38].

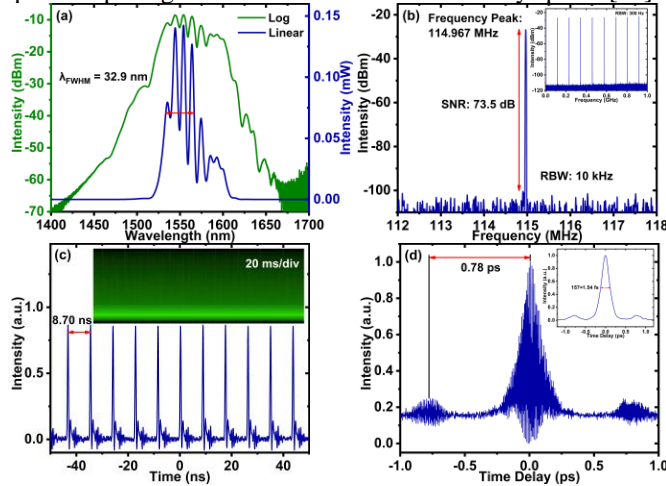

Fig. 5. Characteristics of mode-locked pulses generated by Laser1 at 280-mW pump power. (a) The optical spectrum; (b) The fundamental RF spectrum and its harmonics (inset); (c) The pulse train and oscilloscope screenshot on the large time scale (inset). (d) Fringe-resolved and intensity autocorrelation traces (inset).

Intensity-dependent nonlinear effects such as SPM have a profound impact on the spectral bandwidth and structure of the mode-locked pulses. We have investigated the spectra and autocorrelation traces at different pump powers. The spectral FWHM and shape exhibit specific features, in comparison to dissipative-soliton and similariton fiber lasers [39], [40]. As depicted in Fig. 6(b), when the pump power is below 220 mW, two dominant high-intensity spikes at short wavelengths correspond to the spectral FWHM below 20 nm. Increasing the pump power above 240 mW results in stronger spectral spikes at long wavelengths and a weaker first spectral spike at short wavelengths. Consequently, the spectrum is broadened beyond 30 nm, and the intensity center of the spectrum exhibits a redshift. The apparent broadening of the spectrum results in slight narrowing of the pulse duration from 176 fs to 157 fs, as shown in Fig. 6(c). Besides, the parasitic sidelobes on both sides of the autocorrelation trace become indistinct, indicating enhancement of the pulse quality. According to Figs. 6(b-c), with an increase of the pump power from 200 mW to 280 mW, the time-bandwidth products (TBPs) are computed to be 0.375, 0.364, 0.658, 0.650, and 0.632, respectively. The pulses at low pump powers approach the Fourier transform limit. Trends of the variation of the average output power, peak power, spectral FWHM, and pulse duration of the generated pulses are summarized, as functions of the pump power, in Fig. 6(a). The maximum output power is 7.71 mW, corresponding to the single-pulse energy of 67.1 pJ and peak power of 427 W.

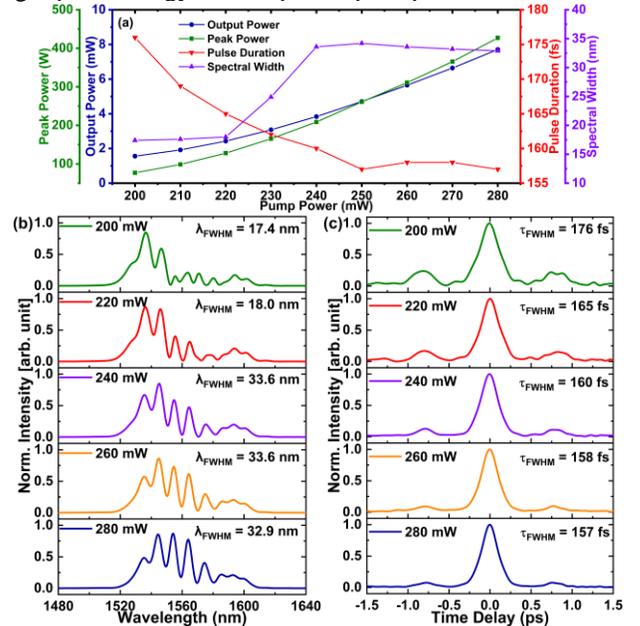

Fig. 6. The effect of the pump power on characteristics of pulses produced by Laser1. (a) The average output power, peak power, pulse duration, and spectral FWHM as functions of pump power; (b) Optical spectra at different pump powers; (c) Autocorrelation traces at different pump powers.

Well-designed fiber lasers delivering stretched pulses should feature low timing jitter, which strongly depends on the pulse formation mechanism [41]. Furthermore, all-PM fiber oscillators are highly sought for constructing low-noise optical frequency combs [42], [43]. The all-PM fiber laser proposed here combines the advantages of both a robust laser cavity and an advantageous pulse-shaping mechanism. A phase noise analyzer (Rohde & Schwarz, FSWP8) is employed to characterize the phase and amplitude noise performance when the fiber laser operates in the free-running mode, without any noise suppression. The InGaAs photodetector converts a few milliwatts of pulsed light into RF signals that are fed directly into the phase noise analyzer. The phase noise power spectral density (PSD) and the corresponding integrated phase noise are presented in Fig.7(a). In the frequency range from 10 Hz to 2.3 kHz, the phase noise value ranges between -60 and -140 dBc/Hz. The observed spikes are caused by the effect of environmental fluctuations on the laser and the intrinsic noise of the photodetector. The noise spectrum then slowly drops to a minimum value of -167 dBc/Hz, in the frequency range from 2.3 kHz to 400 kHz. The phase noise curve ends up maintaining the lowest value in the high-frequency range. The green curve displays the integration of the phase noise PSD, producing the timing jitter of 284 fs integrated from 100 kHz to 10 MHz. Figure 7(b) shows that the relative intensity noise (RIN) curve gradually decreases at the 1 kHz offset frequency until it reaches -150 dBc/Hz. Multiple peaks in the frequency range from 10 Hz to 1 kHz are attributed to acoustic noise and mechanical vibrations [44]. Integrating the entire frequency range from 10 Hz to 10 MHz yields the integrated RIN of 0.0224%.

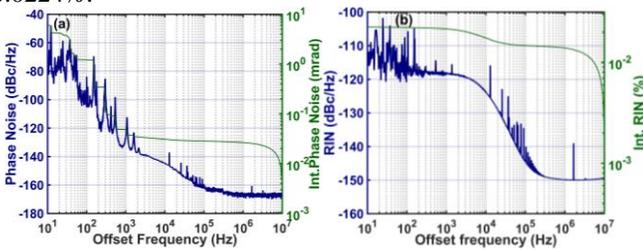

Fig. 7. Noise characteristics of the Laser1 operating in the free-running mode. (a) The phase noise PSD and the corresponding integrated phase noise; (b) The RIN PSD and the corresponding integrated RIN.

### B. Characteristics of Pulses Produced by Mode-locked Laser2

Laser2 is constructed with fewer optics, making this setup simpler and more compact. The compactification of the free-space part is beneficial to reduce the coupling loss of the Semi-WDM and increase the repetition rate of the generated pulses. According to the discussion and analysis in Section 2, the position of the non-reciprocal phase shifter, adopted in the structure of Laser2, significantly changes the modulation depth, phase bias, and modulation period of the reflectivity curve, thereby affecting the evolution of the orthogonally polarized pulses in the laser cavity. Laser2 enters pulsed operation in the mode-locked state at the pump power of 380 mW, which is lower than the self-starting mode-locking threshold of Laser1. Clean single-pulse operation emerges for the pump powers ranging from 170 mW to 250 mW. Figure 8 shows the basic lasing performance of Laser2 at the 250-mW pump power. Compared to Fig. 5(a), the spectrum shown in Fig. 8(a) exhibits similar features, that is, several spectral spikes in the central part and sidebands on both sides of the spectrum. The spectral sidebands may be eliminated by recently developed techniques such as soliton distillation [45], [46]. The difference is that the three main spectral spikes correspond to a wider spectral FWHM of 44.5 nm. Figure 8(b) shows that the measured RF spectrum is centered at the fundamental repetition rate of 116.757 MHz, corresponding to a pulse interval of 8.56 ns [Fig. 8(c)]. An SNR of 78.6 dB is demonstrated, which is higher than the case in Laser1. This is mainly attributed to the simpler cavity structure of Laser2. Accordingly, the generated pulses possess better time-frequency domain properties. The pulse duration, directly measured at the output port, is 90 fs, without any dispersion compensation, as shown in Fig. 8(d). Parasitic sidelobes on both sides of the autocorrelation trace almost disappear, implying better pulse quality than provided by Laser1.

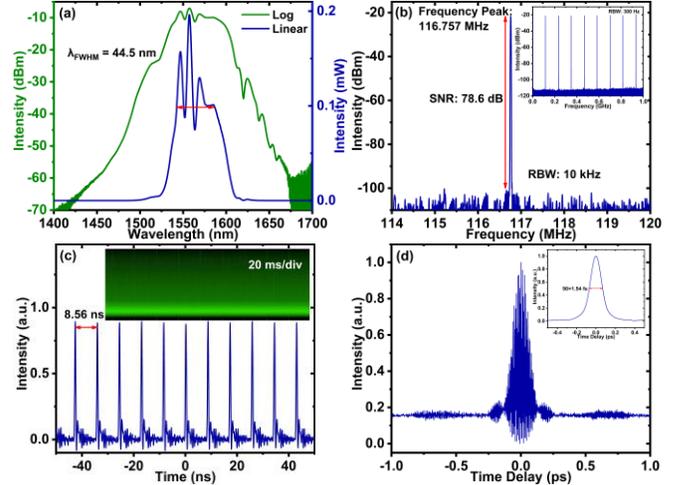

Fig. 8. Characteristics of mode-locked pulses generated by Laser2 at 250-mW pump power. (a) The optical spectrum; (b) The fundamental RF spectrum and its harmonics (inset); (c) The pulse train and oscilloscope screenshot on the large time scale (inset). (d) The fringe-resolved and intensity autocorrelation traces (inset).

Compared with Laser1, the stretched pulse generated by Laser2 exhibits significantly different optical properties, as shown in Fig. 9(a). Varying the pump power from 250 mW to 170 mW corresponds to a linear drop of the average output power from 19.16 mW to 10.75 mW, which is more than twice the average output power of Laser1. Further, in comparison to Laser1, the output pulse of Laser2 characterizes a wider spectral bandwidth and a narrower pulse duration, resulting in a nearly fourfold increase in the peak power. The maximum output power and peak power achieved in Laser2 are 19.16 mW and 1824.2 W, respectively. Figures 9(b-c) show the effects of the pump power on spectral and temporal characteristics of stretched pulses delivered by Laser2. The spectral FWHM is mainly contributed to by three high-intensity spectral spikes. The middle-intensity spike exhibits a blue shift, following the increase of the pump power. At the pump power of 190 mW, Laser2 emits stretched pulses with a maximum spectral FWHM of 50.4 nm and the narrowest pulse duration of 79 fs, which is the shortest for known PM fiber lasers mode-locked by means of NPE. The calculated TBPs are 0.350, 0.487, 0.483, 0.483, 0.490 at pump powers from 170 mW to 250 mW, respectively. The case at 170-mW pump power is close to the Fourier transform limit value of 0.315.

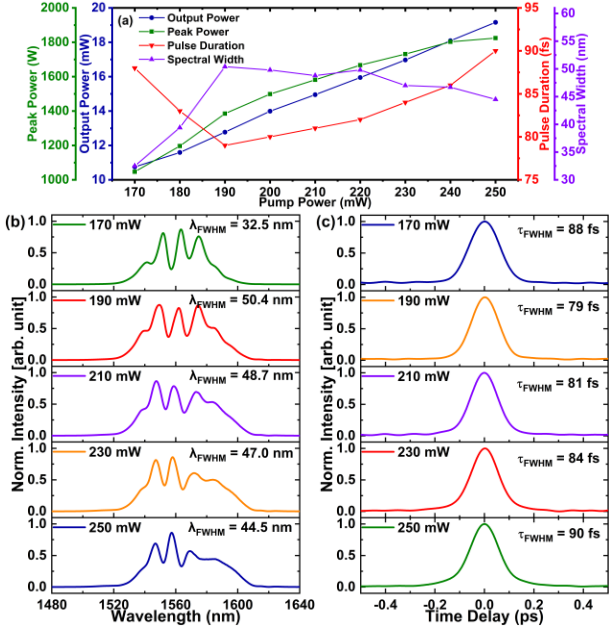

Fig. 9. The effect of the pump power on characteristics of pulses produced by Laser2. (a) The average output power, peak power, pulse duration, and spectral FWHM as functions of pump power; (b) Optical spectra at different pump powers; (c) Autocorrelation traces at different pump powers.

The phase-noise spectrum and the corresponding integration are plotted in Fig. 10(a). The phase-noise curve has characteristics similar to those produced by Laser1. The timing jitter, integrated from 100 kHz to 10 MHz offset frequency, is 209 fs, which is lower than the value produced by Laser1. Figure 10(b) presents the RIN spectrum, which ranges between -110 and -160 dBc/Hz. Overall, the RIN level of Laser2 is lower than that of Laser1, achieving an integrated RIN of 0.0109% in the range of 10 Hz-10 MHz. It has been predicted that short pulse duration and close-to-zero intracavity dispersion can reduce the timing jitter induced by the amplified spontaneous emission (ASE) noise [47], [48]. In dispersion-managed fiber lasers, the generated pulse stretches and compresses as it travels through the laser cavity. The two intracavity positions with the shortest pulse durations are in the middle of the normal- and anomalous-dispersion fiber segments [4]. These two positions correspond to the two ends of the PM fiber of the proposed linear-cavity fiber laser. It is reasonable to conjecture that the pulse extracted at the PBS characterizes a shorter pulse duration than other positions. Therefore, the stability against the noise is more favorable at the PBS than at other positions.

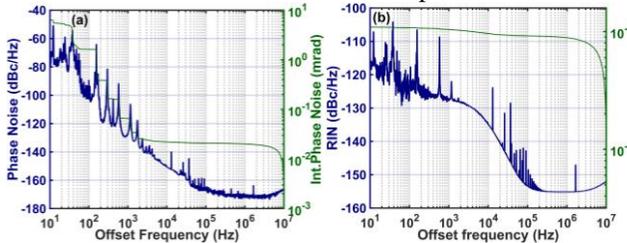

Fig. 10. Noise characteristics of the Laser2 operating in the free-running mode. (a) The phase noise PSD and the corresponding integrated phase noise; (b) The RIN PSD and the corresponding integrated RIN.

The pulse-performance characteristics of Laser1 and Laser2 are compared in Table 2. It is obvious that the characteristics of Laser2 are superior to those of Laser1 in terms of the pulse energy, quality, and noise level. The reason is mainly due to the difference in the evolution of the orthogonally polarized pulses in the two laser setups, which directly leads to the differential pulse selection by the proposed NPE SAs. In addition, Laser2 has a simpler structure that uses one fewer FR, which is beneficial for reducing the intra-cavity loss and increasing the single-pulse energy. Therefore, the peak power of the stretched pulses delivered by Laser2 is much higher than that of Laser1. The stretched pulses produced by Laser2 exhibit better optical properties and negligible pulse sidelobes in autocorrelation traces, which contribute to better noise characteristics. It is natural to consider whether the proposed fiber lasers can achieve a higher pulse repetition rate or a shorter pulse duration. The method to increase the pulse repetition rate is to use high-gain PM fibers or adopt bidirectional pumping to enhance the CW intensity. Achieving a shorter pulse duration requires more precise control of the intracavity dispersion map.

TABLE II
PULSE PERFORMANCE COMPARISON FOR THE TWO PROPOSED SETUPS

| | $F_{rep}$ (MHz) | $P_{pump}$ (mW) | $P_{av}$ (mW) | $E_p$ (pJ) |
|---|---|---|---|---|
| Laser1 | 114.97 | 280 | 7.71 | 67.1 |
| Laser2 | 116.76 | 250 | 19.16 | 164.1 |
| | $\lambda_{FWHM}$ (nm) | $\tau_{FWHM}$ (fs) | Integrated PN (fs) [100 kHz-10 MHz] | Integrated RIN [10 Hz-10 MHz] |
| Laser1 | 32.9 | 157 | 284 | 0.0224% |
| Laser2 | 44.5 | 90 | 209 | 0.0109% |

$F_{rep}$, repetition rate; $P_{pump}$, pump power; $\lambda_{FWHM}$, spectral FWHM; $P_{av}$, average output power; $E_p$, single-pulse energy; $\tau_{FWHM}$, pulse duration; PN, phase noise; RIN, relatively intensity noise.

## IV. CONCLUSIONS

We have demonstrated the operation of two all-PM linear cavity fiber-laser setups (Laser1 and Laser2), which operate in the mode-locked regime maintained by two novel NPE SAs (saturable absorbers) with a simple structure. The SA reflectivity as a function of the nonlinear phase shift is theoretically analyzed to investigate the pulse evolution and mode-locking mechanism. Both setups generate stretched pulses, in the parameter region which supports accurately adjusted dispersion engineering. At the pump power of 280 mW, Laser1 generates ultrashort pulses with temporal duration 157 fs and energy 67.1 pJ, at the repetition rate of 114.97 MHz. The stretched pulses delivered by Laser2 exhibit a wider spectral bandwidth, shorter pulse duration, and higher single-pulse energy. To the best of our knowledge, the shortest pulse duration 79 fs, produced by Laser2, is the smallest value provided by PM fiber lasers mode-locked by means of NPE. The achieved superior optical characteristics, along with the environmental stability, make the proposed seed oscillators highly attractive for many applications, such as optical frequency comb generation, ranging, and imaging.

**Xuanyi Liu** (Student Member, IEEE, Student Member, Optica) received the B.S. degree in the College of Electronic Science and Engineering from Jilin University, Changchun, China, in 2017, the Master degree in the School of Electronic and Computer Engineering, Peking University Peking University, Shenzhen, China, in 2020. He is currently pursuing the PhD degree with the Tsinghua Shenzhen International Graduate School and Tsinghua-Berkeley Shenzhen Institute, Tsinghua University.

**Feng Ye** (Student Member, IEEE, Student Member, Optica) received the B.S. degree in photoelectric information science and engineering from South China Normal University, Guangzhou, China, in 2019, the Master degree in the School of Electronic and Computer Engineering, Peking University, Shenzhen, China, in 2022. He is currently pursuing the PhD of Science degree with the School of Electronic and Computer Engineering, Peking University.

**Minghe Zhao** (Student Member, IEEE, Student Member, Optica) received the B.S. degree in the school of optoelectronic science and engineering from University of Electronic Science and Technology of China, Sichuan, China, in 2020. He is currently pursuing the PhD of Science degree with the School of Electronic and Computer Engineering, Peking University.

**Boris A. Malomed** (Senior Member, Optica) received the Ph.D. degree from the Moscow Physico-Technical Institute, Russia, in 1981, and the Doctor's of Science degree (habilitation) from the Institute for Theoretical Physics of the Academy of Sciences of Ukraine (Kiev) in 1989. He has been working at the Tel Aviv University (currently, as a Professor with chair "Optical Solitons") since 1991. He was a divisional associate editor of Physical Review Letters (responsible for the area of "laser physics") in 2009–2015. Currently he is an editor of Phys. Lett. A, Chaos, Solitons & Fractals, and Frontiers in Physics, and an editorial board member of Journal of Optics, Scientific Reports, Optics Communications, Photonics, and Chaos. His research interests include nonlinear optics, Bose–Einstein condensates and matter waves, pattern formation in nonlinear dissipative media, dynamics of nonlinear lattices, etc. His current h-index is 83 (as per Web of Science and Scopus) and 95 (as per Google Scholar).

**H. Y. Fu** (Senior Member, IEEE, Life Member, Senior Member, Optica) is currently a tenured-associate professor of Tsinghua Shenzhen International Graduate School (SIGS), Tsinghua University. He received the B.S. degree in electronic and information engineering from Zhejiang University, Hangzhou, China, and the M.S. degree in electrical engineering with specialty in photonics from Royal Institute of Technology, Stockholm, Sweden, and the Ph.D. degree from the Department of Electrical Engineering from Hong Kong Polytechnic University. His research interests include integrated photonics and its related applications, fiber optical communications, fiber optical sensing technologies. From 2005 to 2010, he was a research assistant and then research associate with Photonic Research Center, the Hong Kong Polytechnic University. From 2010 to April 2017, he was a founding member and leading the advanced fiber optic communications research at Central Research Institute, Huawei. He was a project manager of All-Optical Networks (AON), which was evolved to a company-wide flagship research project that covers whole aspects of next generation optical communication technologies to guarantee Huawei's leading position. He was also a representative for Huawei at several industry/academic standards/forums. He was an active contributor at IEEE 802.3 Ethernet and Optical Internetworking Forum (OIF) where he was an OIF Speaker from 2012 to 2013. Dr. Fu is senior member of IEEE and life member of OSA, SPIE. From 2017, he is an advisor of OSA Student Chapter at TBSI, Tsinghua University. From 2020, he is advisor of IEEE Photonics Society Student Branch Chapter and SPIE Student Chapter at Tsinghua SIGS. He has authored/coauthored more than 280 journal or conference papers, 3 book chapters, over 80 grant/pending China /Europe/Japan/ US patents.

**Qian Li** (Senior Member, IEEE, Senior Member, Optica) received the Bachelor of Science degree from Zhejiang University, Hangzhou, China, in 2003, the Master of Science degree from the Royal Institute of Technology (KTH), Stockholm, Sweden, in 2005, and the Ph.D. degree from the Hong Kong Polytechnic University, Hong Kong, in 2009. After graduation she was a Visiting Scholar at the University of Washington, Seattle and Postdoctoral Fellow at the Hong Kong Polytechnic University. In 2012 she joined School of Electronic and Computer Engineering (ECE) in Peking University as an Assistant professor. Since 2013 she is Associate Professor at ECE. Her research interests include nonlinear optics, ultrafast optics and integrated optics. Dr. Li is members of Institute of Electrical and Electronics Engineers (IEEE) and senior member of the Optical Society of America (OSA). From March 2017 to April 2019, she is Vice Chair of IEEE ED/SSC Beijing Section (Shenzhen) Chapter and Chair for EDS. From 2015 she is an advisor of OSA Student Chapter in Peking University Shenzhen Graduate School. From 2019 she is an advisor of Peking University Shenzhen Graduate School IEEE Photonics Society Student Branch Chapter.